Ryosuke Nakayama, Sohei Saito, Takuo Tanaka[*], and Wakana Kubo[*]

# Metasurface absorber enhanced thermoelectric conversion

**Abstract:** Metasurfaces are artificial thin materials that achieve optical thickness through thin geometrical structure. This feature of metasurfaces results in unprecedented benefits for enhancing the performance of optoelectronic devices. In this study, we report that this metasurface feature is also essential to drive photo-thermoelectric conversion, which requires the accumulation of thermal energy and effective heat conduction. For example, a metasurface-attached thermoelectric device placed in an environment with uniform thermal radiation generates an output voltage by gathering the thermal energies existing in the environment and creating an additional thermal gradient across the thermoelectric element. In contrast, when a 100-μm-thick-carbon-black-coated electrode was used instead of the metasurface, the device showed lower thermoelectric performance than that of the metasurface-attached device although carbon black exhibits higher infrared absorption than the metasurface. These results indicate that metasurface characteristics of optical thickness and thin geometrical structure for achieving the high thermal conductance are essential in enhancing the performance of photo-thermoelectric devices in terms of the effective collection of thermal energies and conduction of local heating.

## 1 Introduction

Metasurface absorbers (MAs) are artificial absorptive materials that consist of sub-wavelength artificial structures. By carefully engineering their structures, MAs can realize strong light absorption despite being extremely thin. Unlike natural materials, MAs can simultaneously satisfy both large optical thickness and thin geometrical structure. Landy et al. reported a perfect metamaterial absorber consisting of a thin metal layer and a structure sandwiching a thin dielectric layer [1]. It exhibited approximately 100% absorption, whereas the geometrical thickness of the metamaterial was smaller than the incident light $(\lambda_0)/35$. In fact, the highly absorptive property of MAs can be applied to enhance the performance of optoelectronic devices such as solar cells and photodetectors because these devices require strong absorption to gather photons and thin geometrical structure to allow the transport of carriers without changing their migration lengths. Considering the properties of this metamaterial, Atwater and Polman proposed using plasmonic metamaterials to enhance the performance of solar cells [2]. Similar optoelectronic devices have been realized using hot carriers generated in plasmonic metamaterials. Photodetectors driven by hot-carrier injections require thinner plasmonic metamaterials that can facilitate the injection of hot carriers whose lifetimes are on the order of nanoseconds [3-9]. Further, these results support the hypothesis that the metasurface characteristics of optical thickness and thin geometrical structure are beneficial for enhancing the performance of optoelectronic devices. However, these optoelectronic devices primarily function in the visible light region [10], and only a few studies report enhancement in the performance of optoelectronic devices driven in the infrared (IR) region [11-13].

Previously, we demonstrated metamaterial thermoelectric conversion, a novel approach, using an optoelectronic device driven in the IR region that can produce electricity even in an environment with a uniform thermal radiation [14,15]. This device enables us to utilize thermal energy in the surrounding environment, and its driving mechanism is based on plasmonic photo-thermoelectric conversion [16-23]. When an IR MA-fabricated Cu electrode is attached to one end of a thermoelectric device, the MA absorbs the thermal radiation emitted from the surrounding media and converts the absorbed energy to plasmonic local heat. When the MA-attached thermoelectric device is placed in an environment with a uniform thermal radiation, the MA absorbs thermal radiation and generates plasmonic local heating because of absorption loss. The plasmonic local heating propagates to the thermoelectric element via the Cu electrode, which creates an additional thermal gradient across the thermoelectric element and thermoelectric power generation occurs in an environment with a uniform thermal radiation.

The critical factors of such metamaterial thermoelectric conversion include gathering thermal energy from the surrounding media and conduction of plasmonic local heat to a thermoelectric element, which can be achieved by its optical thickness and thin geometrical structure that makes the local heat conduct to the Cu electrode. However, in our previous investigation, we could not answer two questions.

1. How important is the thin geometrical structure of an absorber that drives the metamaterial thermoelectric conversion?

2. Can any other absorber drive thermoelectric conversion in an environment with uniform thermal radiation?

To examine whether other effective absorbers can drive thermoelectric conversion in an environment with a uniform thermal radiation and clarify how metasurface features contribute to thermoelectric conversion, we compared the thermoelectric performance of devices loaded with the MA and a carbon black (CB) layer, which is a more effective IR absorber. Here, MA is a material that satisfies the optical thickness and thin geometrical structure requirements. In contrast, CB only meets the requirement for optical thickness, but not thin geometrical structure. By comparing the thermoelectric performance of the devices loaded with these materials, we can discuss the significance of the metasurface characteristics in the enhancement of optoelectronic devices driven in the IR region. This information will be valuable for developing guidelines to achieve highly efficient IR optoelectronic devices.

As metamaterial thermoelectric conversion allows the arbitrary control of thermal distribution, it can aid in developing better energy harvesting devices that can utilize thermal energy regardless of the thermal distribution. However, examining optimal absorber is essential for further improving the thermoelectric conversion efficiency. Therefore, we investigated the thermoelectric conversion performance of MA and CB in an environment with uniform thermal radiation.

## 2 Materials and methods

A MA consisting of a Ag film, calcium fluoride ($CaF_2$) layer, and Ag disk (diameter: 1.74 μm) was fabricated on a Cu electrode whose width, length, and thickness are 4 mm, 6 mm, and 300 μm, respectively, using electron-beam lithography and thermal evaporation (Figure 1A). The periodicity of the Ag disk array was 3.0 μm. The MA pattern area was 2.1 × 2.1 $mm^2$ and located at the center of the Cu electrode. The film thicknesses of the layers were 150 nm, 60 nm, and 100 nm. The total thickness of the MA was 310 nm, indicating that the MA has an ultrathin structure. Figure 1B depicts the top-view surface morphology of the MA. We confirmed that the MA was successfully fabricated as we designed.

The MA-fabricated electrode was attached to one end of a p-type bismuth antimony telluride thermoelectric element ($Bi_{0.3}Sb_{1.7}Te_3$) provided by Toshima Manufacturing Co., Ltd. using a soldering paste. The $Bi_{0.3}Sb_{1.7}Te_3$ element has a cross-sectional area of 1 × 2 $mm^2$ and a length of 7.6 mm, respectively. The measured Seebeck coefficient of the element was –140 μV/K. A control electrode consisting of Ag (150 nm) and $CaF_2$ (60 nm) layers deposited on a Cu electrode was attached to the other end of the $Bi_{0.3}Sb_{1.7}Te_3$ electrode (Figure 1A, left side). The difference between the MA and control electrodes is only the Ag disk. The control electrode does not have significant IR absorption in the wavelength range from 2–24 μm (Figure 2(A), dashed black line). We refer to this device as a MA device. In contrast, we prepared a control device that loaded the control electrodes at both ends. It should be noted that the control electrode and control device are comparisons for the MA electrode and the MA device, respectively. The detailed procedures for preparing the metasurface and controlling thermoelectric elements have been described in our previous study [15].

To create a CB-coated Cu electrode, we coated a CB layer using a CB spray (TA410KS, Ichinen Tasco Co., Ltd.) with an emissivity of 0.94 on a Cu electrode with an arbitrary thickness (Figure 1C, right side). The CB-coated area corresponds to the Cu electrode area because we coated CB on the whole Cu electrode (4 × 6 $mm^2$), which is considerably larger than the MA area (2.1 × 2.1 $mm^2$). The CB electrode was annealed at 100 °C for 15 min after being coated with the CB layer to evaporate residual solvents. Figure 1D shows the top-view image of the CB surface. The thickness of the CB layer was measured using a scanning electron microscope (SEM; SU-8000, Hitachi High-Technologies Corporation). Table S2 lists the estimated and measured film thicknesses of the CB layers.

We prepared a CB device by replacing the MA electrode of the MA device with a CB-coated electrode (Figure 1C). This procedure enabled us to discuss the effect of the electrode on the thermoelectric performance because the $Bi_{0.3}Sb_{1.7}Te_3$ thermoelectric element and the control electrode at the other end were retained.

Each device was fixed on the center of a printed circuit board via a Kapton double-side tape; the substrate was placed in the chamber of an electric furnace (FUL210FA, Advantec). The printed circuit board is fixed 5 mm higher than the bottom of the electric furnace by legs attached to the four corners of the substrate to avoid direct heat conduction from the bottom floor. The output voltage across the device was measured using a multimeter (DMM-6500, Keithley Instruments). Both a printed circuit board and a double-side Kapton tape are made of polymers, resulting in their thermal conductivities being around 0.2 W/(m·K). This fact indicates that the printed circuit board and Kapton tape act as a thermal insulator. Gold wires (diameter = 0.1 mm) were connected to both electrodes by the Ag paste to create an electrical connection. The gold wires were connected to Cu wires (diameters = 2 mm), and the Cu wires were connected to the multimeter, which was placed outside the furnace. To prevent any impact on the thermoelectric performance, we ensured that the length of the gold and Cu wires connected to both electrodes are the same. Figure 1(E,

F) presents the experimental setup for the thermoelectric measurement. The nonuniformity in device structure, arrangement, and wiring can affect the thermoelectric properties. Therefore, we paid close attention to the device arrangement and length of the wiring in our experiment. We concluded that our control over the device arrangement and wiring was appropriate because the control device did not exhibit significant thermoelectric properties in an environment with uniform thermal radiation, as shown in Figure S1. This result implies that our experimental setup including a control thermoelectric device, a carbon pod, an electric furnace, and any other materials does not generate an output voltage.

The two calibrated thermistors (Micro-BetaCHIP thermistor probe, Measurement Specialties, Inc., length: 3.3 mm, width: 0.3 mm, wire diameter: 0.15 mm) were placed 2 mm from both electrode surfaces to monitor the environmental temperature near both electrodes. The temperature monitored by these thermistors was defined as the environmental temperature. The convection created by the winds of the furnace fan affects thermal gradient across the thermoelectric element, which influences thermoelectric performance. To eliminate any impact on thermoelectric performance, the entire device was capped with a carbon pod (diameter = 2.5 cm, height = 2.5 cm). These experimental procedures were the same as those reported in our previous study [15].

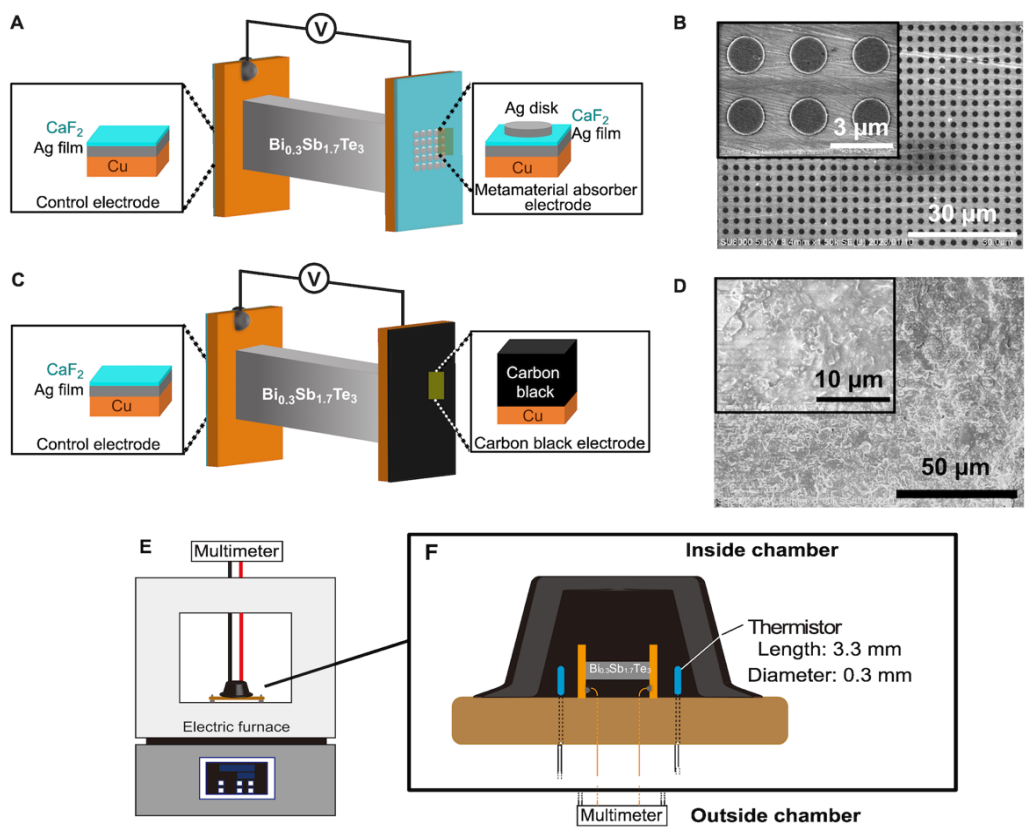

Figure 1: (A) Schematic of the $Bi_{0.3}Sb_{1.7}Te_3$ thermoelectric device loaded with the MA electrode on the right side of the thermoelectric element and (B) top-view SEM images of the MA electrode. (C) Schematic of the $Bi_{0.3}Sb_{1.7}Te_3$ thermoelectric device loaded with the CB electrode on the right side of the thermoelectric element and (D) top-view SEM images of the CB electrode. A control electrode consisting of Ag and $CaF_2$ layers deposited on a Cu electrode was attached to the left side of the thermoelectric element. The scale bars are 3 μm and 30 μm in (B) and 10 μm and 50 μm in (D), respectively. (E, F) Experimental setup for thermoelectric measurement in an environment with uniform thermal radiation. Schematics of (E) an experimental setup using an electric furnace and the (F) settings of the thermoelectric device and thermistors for monitoring environment temperature. Thermoelectric device and thermistors were capped with a carbon pod for eliminating convections created by furnace fan winds.

# 3 Results

Figure 2A compares the absorption spectra of the MA, control, and CB electrodes. The thickness of the CB layer is 100 μm. The MA showed an absorption peak at 6.7 μm, whereas the control electrode indicated by a dashed line did not show significant IR absorption in this wavelength region. The CB layer showed a broader and stronger absorption than the MA in the IR range. The blue line represents the blackbody radiation spectrum calculated at 364 K, which indicates that the CB layer absorbs more thermal radiation than the MA. Figure 2B compares the absorption spectra of the CB layers with thicknesses of 10 μm, 40 μm, 60 μm, and 100 μm, which indicates that every CB layer showed stronger and broader IR absorption than the MA.

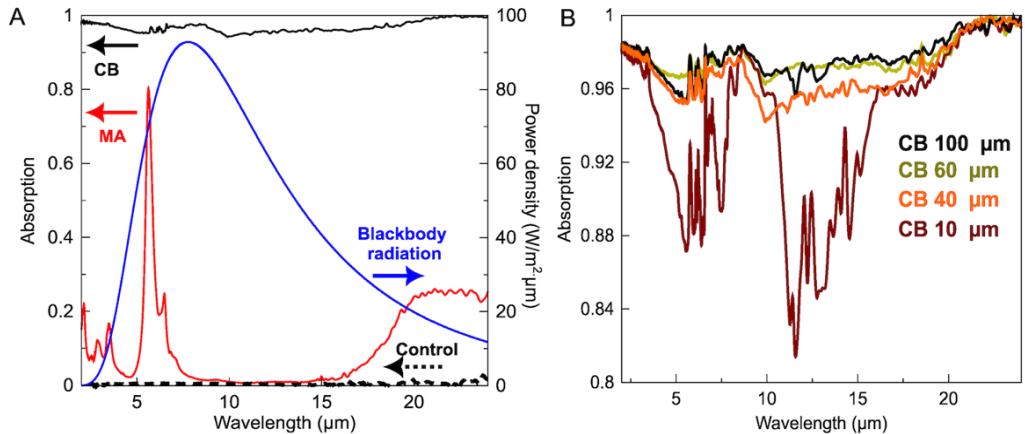

Figure 2: (A) Absorption spectra of the MA (red line), control (black dashed line), and 100-μm-thick CB (solid black line) electrodes. The blue line represents a blackbody radiation spectrum calculated at 364 K. (B) Enlarged absorption spectra of CB layers with thicknesses of 10 μm, 40 μm, 60 μm, and 100 μm, respectively.

Figure 3A shows the dependence of the output voltages of the MA device (red plots) on the measured environmental temperature. The MA device exhibited an output voltage of 19.0 ± 0.1 μV (measured number n = 7) at an environmental temperature of 364 K, which was remarkably larger than that generated on the control device, −1.20 ± 4.4 μV (Figure S1). The error bars in the y-axis in Figure 3A indicate a standard deviation of the measured output voltages. It should be noted that the output voltages at each environmental temperature were measured 4 times at least, meaning that these output voltages were highly reproducible. On the other hand, the control device did not show significant thermoelectric properties at all. This indicates that the output voltage generation is attributed to the thermal radiation absorption of the MA. In addition, we conducted several control experiments that further clarify the contribution of the MA to the output voltage generation.

This result suggests that the MA electrode induced an additional thermal gradient across the $Bi_{0.3}Sb_{1.7}Te_3$ thermoelectric element by thermal radiation absorption, resulting in the generation of an output voltage. Based on the measured Seebeck coefficient of the $Bi_{0.3}Sb_{1.7}Te_3$ element (−140 μV/K), the MA induced a temperature difference of 0.14 K between the ends of the thermoelectric element. In contrast, the 100-μm-thick CB device generated an output voltage of 9.0 ± 4.6 μV at the environmental temperature of 364 K, which was smaller than that generated on the MA device. We measured output voltages generated on these two devices at different environmental temperatures ranging from 320 K to 403 K. Consequently, the MA device exhibited higher output voltages at every environmental temperature than those generated on the 100-μm-thick CB device (Figure 3A). The output voltages generated on the MA device increased linearly with an increase in environmental temperature. In contrast, the output voltages generated on CB devices showed no temperature dependence. A similar pattern was observed for the thermoelectric performance of CB devices with various thicknesses (Figure S2). The increase in output voltage observed for the MA device was attributed to the thermal radiation absorption by MA. The MA generates plasmonic local heat via thermal radiation absorption. The plasmonic local heat propagates to the thermoelectric element via the Cu electrode, which creates an additional thermal gradient across the thermoelectric element.

Figure 3B shows the correlation between output voltages generated on the CB device and the thicknesses of CB layers measured at an environmental temperature of 364 K. The output voltage generated on the MA device is added to the graph for comparison. This graph indicates that the MA device exhibited the highest output voltage among devices loaded with CB electrodes of various CB layer thicknesses. This indicates that CB devices could not generate significant output voltages regardless of the CB layer thickness and absorption amount.

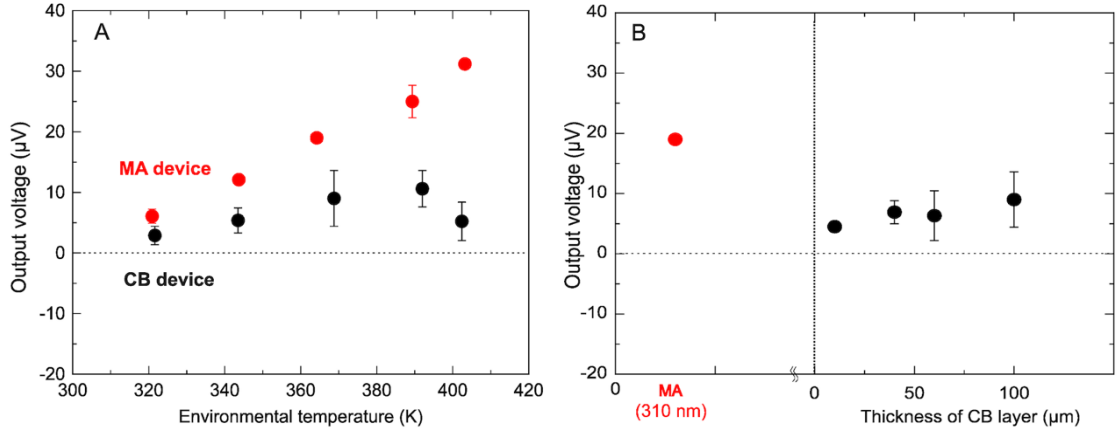

Figure 3: (A) Dependence of output voltages generated on the MA device (red) and a 100 μm-CB device (black) on the measured environmental temperature. (B) Dependence of output voltages on the film thicknesses of CB layers measured at the environmental temperature of 364 K. The output voltage generated on the MA device was shown as a reference.

To determine why the MA device showed higher thermoelectric performance than the CB device, we conducted numerical simulations using COMSOL Multiphysics software and evaluated the heat power absorbed by the MA and CB layers. Figures S3A, B show the measured and calculated absorption spectra of the MA and 100-μm-thick CB layer, respectively. The calculated absorption spectra showed good consistency with those measured, which indicates that our numerical calculation technique was appropriate for estimating the heat power absorbed by these materials.

Table 1 lists the heat power absorbed by the MA and CB layers calculated using the calculated heat power density and volume of each material. The heat power densities of the MA and CB layer were $2.8 \times 10^8$ W/m$^3$ and $2.4 \times 10^7$ W/m$^3$, respectively. Conversely, the total heat powers absorbed by the MA and CB layer were $3.9 \times 10^{-4}$ W and $5.8 \times 10^{-3}$ W, respectively, which indicates that the CB layer absorbed a larger amount of thermal radiation than that absorbed by the MA because of its broadband absorption property. In contrast, the MA showed higher output voltages than the CB device, which means that the heat power absorbed by the material is not the only factor determining the thermoelectric performance driven in an environment with a uniform thermal radiation.

We considered that not only the absorbed heat power but also the thermal conductivity of each material may help determine the thermoelectric performance because thermal conductivity is responsible for conducting the local heat on the Cu electrode. Thermal conductivities of these materials were measured using the 3ω method, and the results are listed in Table 1. We deposited the CB layer on a nonalkali glass substrate with the same thickness as that of the electrode samples for measuring thermal conductivity. An Al electrode with a width, length, and thickness of 0.25 mm, 1.2 mm, and 100 nm, respectively, was formed on the sample surface by thermal evaporation. Further, we deposited 100-nm-thick CaF$_2$ and 150-nm-thick Ag layers on a glass substrate to prepare a sample imitating the MA. The 3ω method was performed according to previous reports [24-26].

The measured thermal conductivity of the sample imitating MA is 240 ± 50 W/(m·K), which is higher than that of CB (61.8 ± 16 W/(m·K)), indicating that MA can effectively propagate plasmonic local heat to increase the temperature of the Cu electrode. In other words, the measured thermal conductivity of the CB layer indicates that local heat generated by the thermal radiation absorption of the CB layer is unlikely to propagate to the Cu electrode. From the point of view of thermal conductivity, the material with a higher thermal conductivity showed higher output voltages, as shown in Figure 3B. This indicates that the thermal conductivity of the material plays an essential role in thermoelectric conversion driven by an environment with a uniform thermal radiation in terms of conductive local heat propagation.

Based on these characteristics, we hypothesized the mechanisms of the thermoelectric conversion driven by the MA and CB electrodes. The MA electrode absorbs less thermal radiation than the CB electrode because of its narrow band absorption; however, the MA has an ultrathin structure with high thermal conductivity, which leads to effective conductive local heat propagation for heating the Cu electrode. In contrast, CB has a higher heat power because of its broader absorption in the IR region. However, sufficient local heat is not conducted to the Cu electrode because of the lower thermal conductivity and larger film thickness of the CB layer. Therefore, it is expected that the amount of thermal radiation emitted from the CB surface may increase because of local heat retention at its surface.

We calculated the thermal resistance ($R_{th}$) to quantify the local heat transfer across the perpendicular direction of the MA and CB electrode and listed in Table 1. We defined the thermal resistance as

$$R_{th} = \frac{t}{\kappa}, \tag{1}$$

where $t$ and $\kappa$ represent the thickness and measured thermal conductivity of the MA and CB layer, respectively. The calculated thermal resistance of the MA and CB were $1.3 \times 10^{-9}$ and $1.6 \times 10^{-6}$ (K/W), respectively, which indicates that the MA had a higher thermal conductance property compared to that of the CB electrode.

Table 1: Comparison of the geometrical thickness, calculated heat power, measured thermal conductivity, calculated thermal resistance, and output voltages of the MA and CB.

| Material | Thickness (μm) | Calculated heat power (W) | Calculated heat power density (W/m$^3$) | Measured thermal conductivity (W/(m·K)) | Thermal conductivity (reported in the literature) (W/(m·K)) | Thermal resistance (K/W) | Output voltage measured at 364 K (μV) |
|---|---|---|---|---|---|---|---|
| MA | 0.31 | 3.9 x 10$^{-4}$ | 2.8 x 10$^8$ | 240 ± 50 | 430 [27,28] | 1.29 x 10$^{-9}$ | 19.0 ± 0.1 |
| CB | 100 | 5.8 x 10$^{-3}$ | 2.4 x 10$^7$ | 62 ± 16 | 6–174 [29] | 1.62 x 10$^{-6}$ | 9.0 ± 4.6 |
|  | 60 | 3.4 x 10$^{-3}$ | - | 44 ± 4 | - | 9.05 x 10$^{-7}$ | 6.3 ± 4.1 |
|  | 40 | 2.3 x 10$^{-2}$ | - | - | - | - | 6.8 ± 1.9 |
|  | 10 | 5.8 x 10$^{-4}$ | - | - | - | - | 4.5 ± 8.8 |

As proof of this concept, we measured the temperature at the rear side of the electrodes at an environmental temperature of 364 K. We monitored the temperature at the rear side of both electrodes simultaneously because the temperature difference between the right and left electrodes is used to determine the thermoelectric output voltage. Figure S4 shows the experimental setup for measuring the temperatures at the rear side of the electrodes. In this experimental setup, the thermoelectric element was attached to the higher part of the electrode to free space for attaching thermistors. Figure S4(B) presents the arrangement of the Bi$_{0.3}$Sb$_{1.7}$Te$_3$ thermoelectric element and a thermistor on the rear side of the Cu electrode; the thermistor was sandwiched between a Kapton double-side tape to be attached on the rear side of the Cu electrode. We evaluated the temperature at the rear side of electrodes after the environmental temperature reached stable. The thermistors were calibrated according to the method described in Supplementary text S1. After careful calibration, thermistors 1 and 2 exhibited a temperature difference of around 0.015 K when they experienced the same environmental temperature.

Supplementary text S2 describes the procedure to measure the temperature at the rear side of each electrode. Because thermistors have some inaccuracies and fluctuations of their signal in temperature measurement, we measured the temperatures at the rear side of each electrode at least four times to verify its reproducibility. The absolute value of the temperature difference between a MA or CB electrodes and the control electrode on the order of 0.01 K is unreliable; however, the order of the temperature difference between each electrode and the control electrode can be revealed using the same thermistors and measurement procedures. We measured the temperature difference between each electrode and the control electrode twice, and we performed the same measurement using different thermistors and obtained four data sets. Then, we calculated their average. About the detailed procedure, please refer S2.

Table 2 summarizes the measured temperature differences between the MA or CB electrodes and the control electrode. In the case of the MA electrode, the temperature difference corresponded to the difference between the MA and control electrode temperatures. The MA electrode showed the higher temperature difference of 0.06 ± 0.08 K (measurement number: n = 4), whereas the 100-μm-thick CB layer showed 0.00 ± 0.10 K (n = 4). This implies that the MA electrode can induce a larger thermal gradient across the thermoelectric element than in the CB electrode. As mentioned previously, the absolute values of the temperature difference are unreliable because thermistors have some inaccuracies and fluctuations in temperature measurement; however, the order of the temperature difference can be relied on because temperature differences were measured using the same thermistors and temperature measurement procedures. The order of the temperature difference for each electrode was MA > CB, which was consistent with the order of the measured output voltages in an environment with uniform thermal radiation. In contrast, the output voltages estimated from the temperature difference for each electrode were lower than those measured. This could have originated from the low accuracy of the thermistors in measuring temperatures, as mentioned previously.

These results support our hypothesis that optical absorption and high thermal conductance property of the MA are essential for driving the thermoelectric conversion. Moreover, the MA that exhibited the highest output voltages among CB samples with various film thickness was the only material that meets the optical thickness and the high thermal conductance requirements.

Table 2: Temperature difference between the MA or CB electrodes and the control electrodes measured at an environmental temperature of 364 K.

| Material | Temperature difference between each electrode and the control electrode (K) | Output voltage estimated by the temperature difference ($\mu$V) | Measured output voltage ($\mu$V) |
|---|---|---|---|
| MA (310 nm) | 0.06 ± 0.08 (n = 4) | 8.4 | 19.0 ± 0.1 |
| CB (100 $\mu$m) | 0.00 ± 0.10 (n = 4) | 0.0 | 9.0 ± 4.6 |

# 4 Discussion

We examined the thermoelectric performance of the devices loaded with the MA and CB-coated electrodes. We observed that a 100-μm-thick CB layer allowed the highest absorption in the IR region, whereas the device loaded with the MA electrode showed the highest thermoelectric performance among devices. We hypothesized that the optical thickness and high thermal conductance property of the MA are critical for driving thermoelectric conversion in an environment with a uniform thermal radiation. To prove this concept, we measured the temperature at the rear side of each electrode in an environment with a uniform thermal radiation. We found that the order of the temperature difference for each electrode was MA > CB, which was consistent with the order of the measured output voltages in an environment with uniform thermal radiation. These results indicate that metasurfaces with an optical thickness and high thermal conductance property are essential for driving thermoelectric conversion in terms of thermal radiation absorption and effective conductive local heat propagation. Metasurfaces will be the crucial element in the development of highly efficient IR optoelectronic devices. Furthermore, by combining metamaterial thermal engineering techniques such as controlling thermal conduction and radiation, there is a space for improving the device performance of metamaterial thermoelectric conversion in future [30-33].

# Acknowledgement

The authors thank TOSHIMA Manufacturing Co., Ltd. for providing p-type $Bi_{0.3}Sb_{1.7}Te_3$ elements. W. Kubo would like to thank Dr. Yasuyuki Ozeki of The Univ. of Tokyo for fruitful discussion. W. Kubo also thank the Thermal & Electric Energy Technology Foundation (2022-013), the JSPS KAKENHI (Grant number JP-20K05261), and the JSPS Core-to-Core Program. The authors acknowledge financial support from JST CREST (Grant Number JPMJCR1904) and KAKENHI (JP-18H03889), Japan.

# Metamaterial absorber enhanced thermoelectric conversion

Ryosuke Nakayama, Sohei Saito, Takuo Tanaka*, and Wakana Kubo*

Supplementary information

Supplemental text S1, S2
Table S1, S2
Fig. S1-S4

Supplemental text

**S1. Calibration of thermistors**

A thermistor probe (Micro-BetaCHIP thermistor probe, Measurement Specialties, Inc.) was used to measure environmental temperature. The thermistor is a negative temperature coefficient (NTC) type and has a probe head with a length and diameter of 3.3 and 0.3 mm, respectively. The wires (diameter: 0.15 mm) were connected to the head of the thermistor, which was connected to a temperature sensing circuit that produced an output voltage linear to the environmental temperature.

First, we calibrated thermistors because each thermistor has an intrinsic temperature property. Thermistor 1 and 2 were affixed with Kapton tape to be exposed to the same temperature environment, and we were put in a carbon pod put in an electric furnace set to 100 °C for the measurement. We attach a thermocouple to the thermistors to monitor the environmental temperature as a reference and calibrate thermistors.

The measured temperature can be calculated using

$$\frac{1}{T} = \frac{1}{T_0} + \frac{1}{B} ln \frac{R}{R_0} \qquad \text{Eq. S(1)}$$

where $B$, $R_0$, and $R$ represent a beta parameter, resistance of the circuit at the standard temperature ($T_0$, 298.15 K), and resistance, respectively.

Then, we adjusted the B value of thermistor 2 to make its measured temperature as close as possible to the one monitored by thermistor 1. Table S1 shows the correlation between $R_0$, $B$, and a temperature difference of thermistors 1 and 2. Even when the B value was adjusted, thermistor 2 showed a temperature difference of 0.02 K relative to thermistor 1.

Table S1. Correlation between $R_0$, $B$, and a temperature difference between the thermistors 1 and 2.

| Thermistor | Resistance of the circuit $R_0$ (Ω) | Adjusted B value | Temperature difference relative to thermistor 1 (K) |
|---|---|---|---|
| 1 | 10,000 | 3994 | 0 |
| 2 | 10,000 | 3992 | 0.015 |

**S2 Measurement of the temperature at the rear side of each electrode**

We utilized thermistors 1 and 2, which were calibrated according to Supplemental text S1, to measure the temperature at the rear side of each electrode. Since thermistors have always some inaccuracies and fluctuations in temperature measurements, we measured the temperatures at the rear side of each electrode at least four times to minimize the inherent measurement errors and fluctuations. Detailed procedure is as follows. First, we connected thermistor 1 to the rear side of the MA or CB electrode and thermistor 2 to the control electrode, and then measured the difference of their temperature twice. After that, we switched the two thermistors and again measured the temperature difference twice. Finally, we averaged these four data.

We adopted this measurement procedure to measure the temperature at the rear side of each electrode using thermistors 1 and 2. Based on this method, we believe that the order of the temperature difference between the MA or CB electrodes and the control electrode accurately reflects the temperature order of the MA and CB electrodes.

Table S2. Measured film thickness of the CB layer of 10 μm, 30 μm, 60 μm and 100 μm. n indicates the measurement number of the sample.

| Sample name | Estimated film thickness of the CB layer (μm) | Measured film thickness of the CB layer (μm) |
| --- | --- | --- |
| 10 μm CB | 10 | 9.4 ± 0.4 (n = 8) |
| 40 μm CB | 40 | 40.8 ± 0.6 (n = 9) |
| 60 μm CB | 60 | 63.9 ± 0.8 (n = 9) |
| 100 μm CB | 100 | 103 ± 1 (n = 7) |

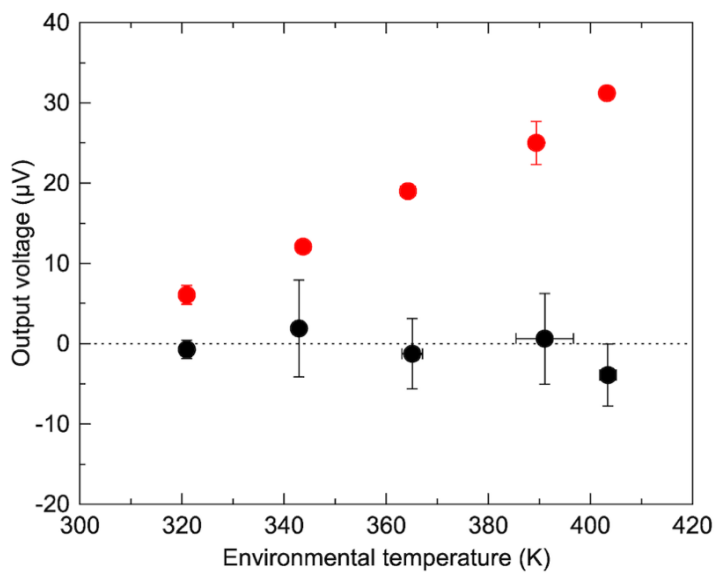

Fig. S1 Dependence of the output voltages generated on the MA device (red) and a control device (black) on the measured environment temperatures.

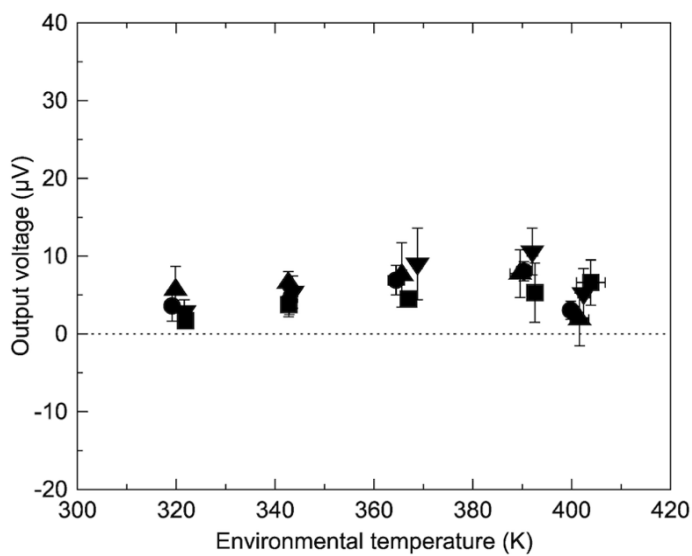

Fig. S2 Dependence of the output voltages generated on the CB device with various thickness of 10 (■), 40 (●), 60 (▲), and 100 (▼) μm on the measured environment temperatures. The error bars in the y-axis indicate the standard deviation of the measured output voltages.

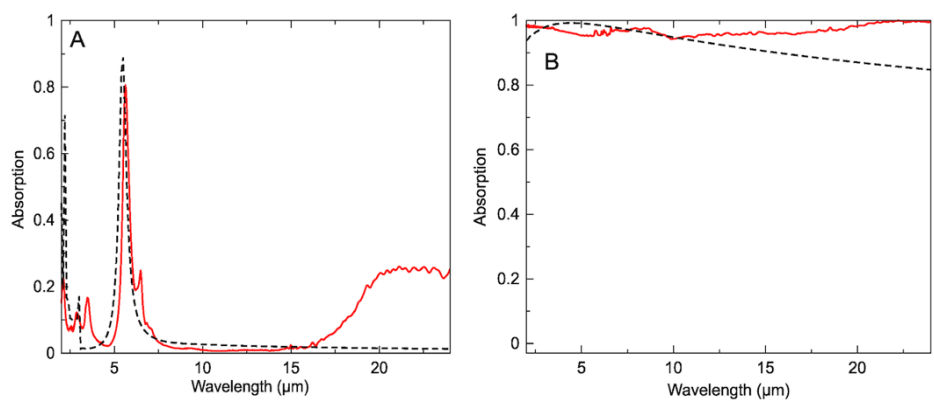

Fig. S3 Measured (red lines) and calculated (dashed lines) absorption spectrum of (A) MA (310 nm) and (B) CB (100 μm) electrodes. The reflection spectra of the samples were measured by microscopic Fourier transform infrared spectrometer (FT/IR-6300, VIRT-3000, JASCO Corporation). The reference spectrum was measured on a bare copper plate. The absorption spectra were calculated subtracting measured reflectance from 1.0.

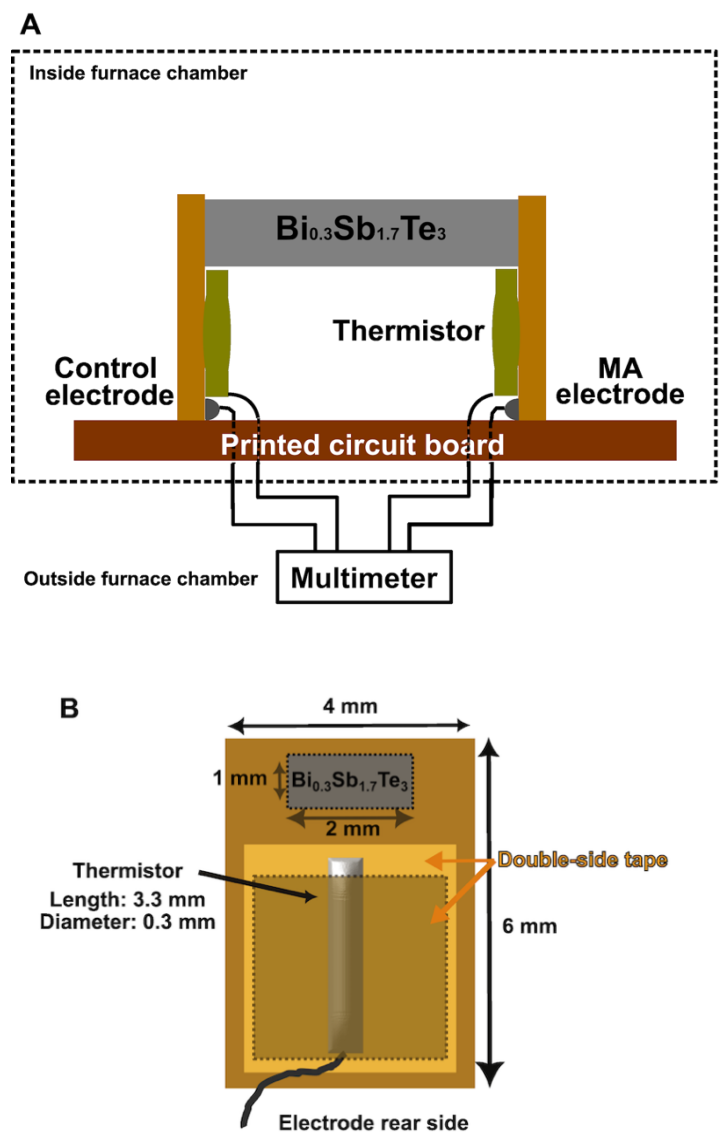

Fig. S4 Experimental setup for measuring the temperatures of the electrode rear side. The MA electrode was exchanged by the CB electrode. Schematics of (A) experimental setup in the X-Z view and (B) arrangements of the $Bi_{0.3}Sb_{1.7}Te_3$ thermoelectric element (1 × 2 mm$^2$) and a thermistor (3.3 mm × 0.3 mm) on the rear side of the Cu electrode. The thermistor was sandwiched between a Kapton double-side tape to be attached on the rear side of the Cu electrode. In the experiment, the thermistor was fully covered by the top Kapton tape.